\documentclass[conference]{IEEEtran}
\IEEEoverridecommandlockouts
\usepackage[letterpaper, top=0.75in, bottom=1.035in, left=0.63in, right=0.625in]{geometry}
\usepackage{hyperref}
\usepackage{cite}
\usepackage{amsmath,amssymb,amsfonts}
\usepackage{graphicx}
\usepackage{textcomp}
\usepackage{tikz}

\def\BibTeX{{\rm B\kern-.05em{\sc i\kern-.025em b}\kern-.08em
    T\kern-.1667em\lower.7ex\hbox{E}\kern-.125emX}}

\usepackage{xcolor}       
\usepackage{enumitem}         
\usepackage{algorithm}        
\usepackage{algpseudocode}    
\ifCLASSOPTIONcompsoc
 \usepackage[caption=false,font=normalsize,labelfont=sf,textfont=sf]{subfig}
\else
 \usepackage[caption=false,font=footnotesize]{subfig}
\fi
\usepackage{orcidlink}
\usepackage{soul,color}
\usepackage[nameinlink,capitalize]{cleveref}
\crefname{equation}{Eq.}{Eqs.}
\crefname{section}{Sec.}{Secs.}
\crefname{figure}{Fig.}{Figs.}

\definecolor{skyblue}{RGB}{135, 206, 235}
\definecolor{lightyellow}{RGB}{255, 255, 128}

\usepackage{makecell}

\begin{document}

\title{
Clone What You Can’t Steal: Black-Box LLM Replication via Logit Leakage and Distillation




}


\author{
\IEEEauthorblockN{Kanchon Gharami \orcidlink{0000-0003-0032-8201}
}
\IEEEauthorblockA{
Department of Electrical Engineering and Computer Science\\
Embry-Riddle Aeronautical University, FL, USA\\
Email: gharamik@my.erau.edu, kanchon2199@gmail.com}
\and
\IEEEauthorblockN{Hansaka Aluvihare \orcidlink{0009-0005-8748-761X}} 
\IEEEauthorblockA{
Department of Mathematics\\
Embry-Riddle Aeronautical University, FL, USA\\
Email: aluvihah@my.erau.edu}
\and
\IEEEauthorblockN{Shafika Showkat Moni \orcidlink{0000-0002-7710-4217}
}
\IEEEauthorblockA{
Department of Electrical Engineering and Computer Science\\
Embry-Riddle Aeronautical University, FL, USA\\
Email: monis@erau.edu}
\and
\IEEEauthorblockN{Berker Peköz \orcidlink{0000-0002-7572-3663}
}
\IEEEauthorblockA{
Department of Electrical Engineering and Computer Science\\
Embry-Riddle Aeronautical University, FL, USA\\
Email: berker.pekoz@erau.edu}
}

\maketitle

\begin{abstract}
Large Language Models (LLMs) are increasingly deployed in mission-critical systems, facilitating tasks such as satellite operations, command-and-control,  military decision support, and cyber defense. Many of these systems are accessed through application programming interfaces (APIs). When such APIs lack robust access controls, they can expose full or top-$k$ logits, creating a significant and often overlooked attack surface. Prior art has mainly focused on reconstructing the output projection layer or distilling surface-level behaviors. However, regenerating a black-box model under tight query constraints remains underexplored. We address that gap by introducing a constrained replication pipeline that transforms partial logit leakage into a functional deployable substitute model clone. Our two-stage approach (i) reconstructs the output projection matrix by collecting top-$k$ logits from under 10k black-box queries via singular value decomposition (SVD) over the logits, then (ii) distills the remaining architecture into compact student models with varying transformer depths, trained on an open source dataset. A 6-layer student recreates 97.6\% of the 6-layer teacher model’s hidden-state geometry, with only a 7.31\% perplexity increase, and a 7.58 Negative Log-Likelihood (NLL). A 4-layer variant achieves 17.1\% faster inference and 18.1\%  parameter reduction with comparable performance. The entire attack completes in under 24 graphics processing unit (GPU) hours and avoids triggering API rate-limit defenses. These results demonstrate how quickly a cost-limited adversary can clone an LLM, underscoring the urgent need for hardened inference APIs and secure on-premise defense deployments.
\end{abstract}


\section{Introduction}
\label{sec:introduction}
Large-language models (LLMs) are rapidly transitioning from research prototypes to operational assets across national security domains. The U.S. Army has fine-tuned transformer-based models, such as The Research and Analysis Center Language Model (TRACLM)~\cite{ruiz2024fine} family, on an 82 million-token doctrine corpus to support planning, training and decision support. These 3 to 7 billion parameter-models, are deployed on secure, in-house graphics processing units (GPUs) to ensure operational continuity and data sovereignty.

Beyond the Army, LLMs are increasingly integrated into critical systems such as command-and-control interfaces, satellite ground station automation, and cyber threat triage platforms~\cite{javaid2024leveraging, zucchelli2025fine, koksal2025milchat, hassanin2025pllm}.  Analysts project that U.S. defense-sector demand for LLM-enabled services will grow from approximately USD \$50 million in 2024 to USD \$1.4 billion by 2030, reflecting a 37.2\% compound annual growth rate~\cite{javaid2024leveraging}. As these models move out of the lab and are exposed via cloud-based application programming interfaces (APIs), their security posture becomes a matter of national resilience. 

Recent research has demonstrated that LLM APIs, particularly those exposing top-$k$ logits, can inadvertently expose internal model structure. For example, fewer than 400 million carefully crafted queries can reconstruct the output projection matrix of GPT-3.5-turbo, costing less than USD \$2,000 at 2025 OpenAI pricing. In comparison, smaller commercial models can be cloned with under two million queries for just around USD \$20~\cite{carlini2024stealing} in 2025. However, these efforts often stop at reconstructing the final layer or mimicking surface-level behavior. They do not yield deployable clones that replicate the target model’s latent reasoning or generalization capabilities under realistic API rate limits. 


This paper addresses that critical gap. We present a practical, two-stage black-box replication pipeline that transforms partial logit leakage into a fully functional substitute model suitable for real-world red-team exercises. Our approach (i) recovers the output projection matrix using fewer than 10,000 top-$k$ logit queries via singular value decomposition (SVD) inspired by Carlini's work~\cite{carlini2024stealing}. We then (ii) distill the remaining architecture into compact student models of varying transformer depths, trained exclusively on open-source dataset.

The contributions of this paper are summarized as follows:

\begin{itemize}
    \item We introduce a two-step \textbf{black-box LLM replication pipeline} using only top-$k$ logits, without access to gradients, weights, or training data.
    
    \item We demonstrate that combining combine partial projection recovery with task-aware distillation \textbf{preserves both output behavior and internal representation geometry}.

    \item Our 6-layer student model replicates \textbf{97.6\% of the teacher model's hidden-state geometry}, with only a 7.31\% perplexity increase. A smaller 4-layer variant retains similar fidelity within a 10.5\% perplexity gap. 
    
    \item Evaluation on unseen test corpus confirms the clones  generalize beyond memorized prompts, \textbf{capturing the teacher model's latent reasoning behavior}.

    \item The full attack completes in under \textbf{24 GPU-hours} and avoids rate-limit defenses using fewer than 10k queries.

    \item We analyze \textbf{trade-offs} between model depth, inference cost, and fidelity to help with defense planning.
\end{itemize}

The rest of the paper is organized as follows.  \cref{sec:related_work} surveys related work in adversarial LLM research.  \cref{sec:background} defines the threat model and background.  \cref{sec:proposed_framework} details our replication framework.  \cref{sec:experimental_analysis} presents experimental results.  \cref{sec:conclusion} concludes with implications and outlines future directions.

\section{Related Works
\label{sec:related_work}}
Research on adversarial attacks against LLMs moved quickly from qualitative conceptual demonstrations to rigorous attacks featuring practical, scalable techniques. This section reviews prior art relevant to model extraction, knowledge distillation, and black-box replication under constrained access.

Carlini \emph{et al.}~\cite{carlini2024stealing} introduced a foundational black-box attack that reconstructs the entire embedding–projection matrix of commercial LLMs such as \textit{gpt‑3.5‑turbo} and \textit{PaLM‑2} by exploiting log‑probability outputs. Their top‑down SVD recovers hidden‑state dimensionality and projection weights with low mean‑square error (MSE), using less than two million queries at a cost of under \$20. The work proves that even heavily guarded APIs can leak non‑trivial internal structure; however, it remains limited to the \emph{final} layer and does not transfer task‑specific reasoning patterns or reduce query cost when only top‑$k$ logits are available.

Panda \emph{et al.}~\cite{panda2024teach} demonstrated that injecting  benign‑looking “poison” sentences during pre‑training can later coerce the model to regurgitate private identifiers seen only once during fine‑tuning. The attack achieves up to 50\% success at leaking 12‑digit secrets. While their three‑phase pipeline (poisoning, fine‑tuning, inference) reveals a severe privacy hole, it assumes write access to the training corpus and does not attempt to copy the victim model; instead it weaponizes memorization.

Liu and Moitra~\cite{liu2024model} propose a polynomial-time algorithm that can provably steal hidden-Markov or low-rank language models using conditional queries. The method solves a series of Kullback–Leibler (KL)-projected convex programs to recover output distributions within $\epsilon$ total variation. While theoretically elegant, it assumes access to strong conditional-query oracles not typically available in LLM APIs and lacks large-scale validation.

Oliynyk \emph{et al.}~\cite{oliynyk2023know} synthesizes more than 100 papers into a taxonomy that encompasses objectives, attacker knowledge, and defense strategies.  
They tabulate substitute‑model, hyper‑parameter, and side‑channel attacks with watermarking, output perturbation, and monitoring defenses. The survey confirms that high‑fidelity, task‑specific cloning of LLMs is still under‑explored, especially under realistic query budgets.

Feng and Tramèr~\cite{feng2024privacy} design \emph{"data traps"}, single‑use weight perturbations that overwrite themselves with a training example during fine‑tuning and remain latent until the model is inspected. The attack reconstructs dozens of finetuning samples from ViT \& BERT checkpoints with minimal accuracy loss. It illustrates a supply‑chain risk orthogonal to query‑based stealing; defenses like random re‑initialization remain ineffective.

Sha \emph{et al.}~\cite{sha2023can} propose \emph{Cont‑Steal}, a black‑box attack that treats target embeddings as positives in a contrastive loss and steals vision encoders with fewer queries than earlier coordinate‑descent methods. While effective on representation learning models, the method relies on surrogate linear evaluation and has not been adapted to autoregressive LLMs.

Guan \emph{et al.}~\cite{guan2024large} used instruction-tuned LLMs to predict hidden links between graph nodes, outperforming basic similarity-based methods. Their prompts adapt to different dataset structures, but the focus was on graph privacy, not model stealing.

Chernyshev et al.~\cite{chernyshev2024forensic} studied how to detect indirect prompt injection attacks by analyzing LLM agent logs. They used the AgentDojo benchmark and tested 12 LLMs on tasks with hidden injections. By comparing model outputs with ground truth, they measured how well each model spotted malicious behavior. Models like Claude and Gemini performed best, but others struggled across domains. The method shows promise for forensic use, though results are still early and limited to synthetic test cases and one type of attack.

Irtiza et al.~\cite{irtiza2024llm} introduced LLM-Sentry, a defense system that filters harmful prompts before they reach the model. It works in two stages: first, it uses a zero-shot classifier with a sliding window to detect intent shifts, and then applies a retrieval module that checks against a growing database of harmful prompts. The system is model-agnostic and doesn’t need retraining when new threats appear. Tested on GPT-3.5, Gemini, and Mistral, it reached 97\% accuracy. The main strengths are its generality and strong detection rate, but it adds extra computation and depends on good translation and prompt coverage.

He et al.~\cite{he2024data} propose a backdoor-based attack that steals private data from customized LLMs without needing access to model internals. Their method works in two stages: during fine-tuning, attackers inject poisoned samples containing a secret trigger; later, they query the model with that trigger to extract sensitive information. The model behaves normally for regular users, making the attack stealthy. Tested on GPT-3.5 and Mistral-7B, the method achieves high success (up to 92.5\%) and outperforms prior work like PLeak. The main downside is the need to tamper with the model during customization, which may not always be possible in practice.

Adaptive dense‑to‑sparse constrained (ADC) optimization by Kai \emph{et al.}~\cite{hu2024efficient} and context‑fusion multi‑turn jailbreaking by Sun \emph{et al.}~\cite{sun2024multi} reduce the cost and detection rate of prompt‑based exploits, attaining state‑of‑the‑art success on HarmBench~\cite{mazeika2024harmbench}. These techniques confirm that alignment can be circumvented, yet they produce attack strings, not substitute models, and still rely on thousands of gradient calls.


While prior work has addressed projection matrix recovery\cite{carlini2024stealing} and representation-level attacks \cite{sha2023can} independently, to our knowledge, no existing method integrates these techniques into a deployable, task-aware cloning pipeline under realistic API constraints. Our framework addresses this underexplored challenge, enabling high-fidelity replication of black-box LLMs using only top-$k$ logit access.

\begin{figure*}[htbp]
  \centering
  \includegraphics[width=1\linewidth]{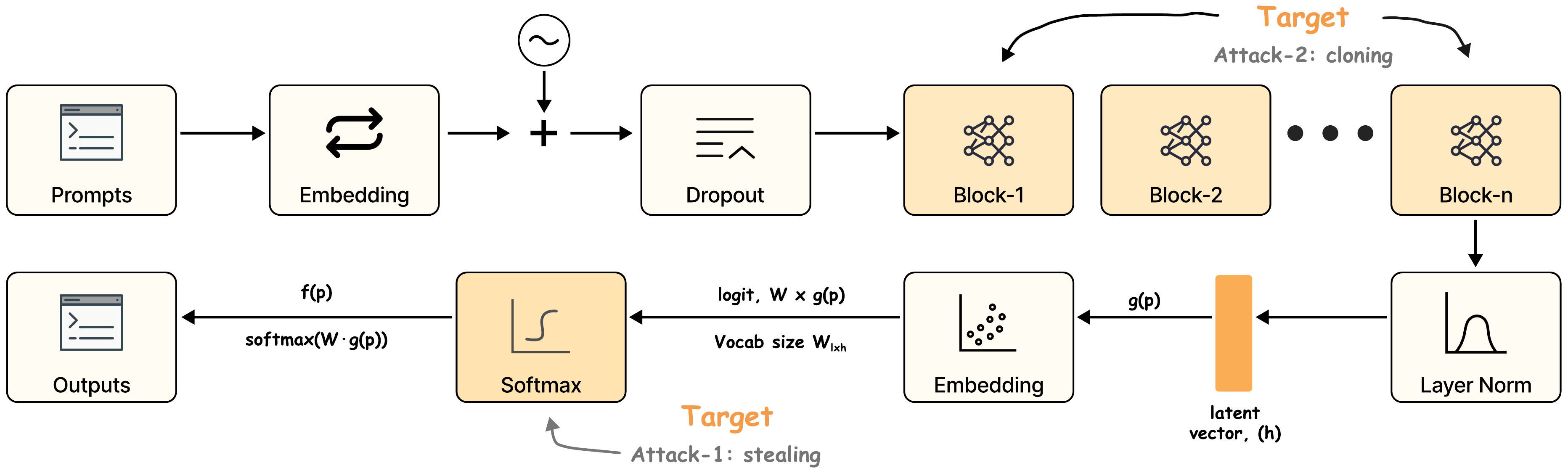}
  \caption{High–level data flow inside a transformer LLM.  \emph{Attack–1} steals the \emph{last–layer projection matrix}~$W$; 
  \emph{Attack–2} distills the full stack of  transformer blocks into a light student model.\label{fig:overview}}
\end{figure*}

\section{Threat Model \& Background
\label{sec:background}}

\subsection{System Under Attack}
We target a transformer-based LLM $\mathcal{F}$ deployed in a black-box
accessible via an API interface, a common configuration in both commercial and
mission-critical environments. Given input prompt $\mathbf{p} = (t_1, \dots, t_N)^\top$, $\mathcal{F}$, through a stack of transformer layers, computes a hidden representation $g(\mathbf{p}) \in \mathbb{R}^d$, where $d$ is the hidden dimension, unknown to the attacker. This hidden state is then projected to vocabulary space via an output projection matrix $\mathbf{W} \in \mathbb{R}^{V \times d}$, where $V$ is the vocabulary size.  The final output distribution is defined as $f_T(\mathbf{p})$:
\begin{equation}
    f_T(\mathbf{p}) = \mathrm{softmax}(\mathbf{W}\,g(\mathbf{p})).
\end{equation}
 Adversary's targets are twofold:
\begin{itemize}[leftmargin=1.5em]
    \item \textbf{Stealing projection:} The output projection matrix $\mathbf{W}$, translating internal representations to token predictions.
    \item \textbf{Cloning behavior:} Internal transformation function $g(\cdot)$, encapsulating target's reasoning and understanding.
\end{itemize}

\subsection{Adversary Goal and Capabilities}
\subsubsection{Goal}
Construct a high-fidelity \textit{clone} (or \textit{student}) of the \textit{target} (or \textit{teacher}) LLM. The clone should replicate both the output behavior and internal representation geometry of the target, enabling red-team simulations or adversarial testing.

\subsubsection{Capabilities}
\begin{itemize}[leftmargin=2.5em]
  \item \textbf{Black-box access:} The attacker submits prompts and receives top-$k$ logits for next-token predictions, but cannot
  access weights, gradients, training data, or architecture. 
  \item \textbf{Communication constraints:} API enforces rate limits and query quotas, simulating realistic deployments.
  \item \textbf{External resources:} The attacker may leverage public datasets (e.g., WikiText) and commodity computing.
\end{itemize}

\subsubsection{Assumptions}
\begin{itemize}[leftmargin=2.5em]
  \item Hidden dimension $d$ is unknown but satisfies $d \ll V$.
  \item The API returns at least top-$k$ logits, where $k \ge d+1$.
  \item Returned logits are unrounded and unperturbed by noise.
\end{itemize}

While access to logits seem unlikely, Carlini et al.~\cite{carlini2024stealing} demonstrated that even commercial APIs can be manipulated into revealing them through prompt engineering. This reinforces the practical relevance and ground of our threat model.

\section{Proposed Framework}
\label{sec:proposed_framework}

Our attack begins by targeting the only part of the LLM exposed through the API: its final projection layer. As shown in \cref{fig:overview}, first we exploit the structure in the top-$k$ logits returned by the API to recover a low-rank approximation of the projection layer. By querying the model with diverse prompts and applying singular value decomposition (SVD) to the collected logits, we estimate the subspace in which the output weights lie. The second stage focuses on what cannot be directly stolen: the transformer's internal blocks. These layers are wrapped in non-linear activations and deeply entangled, making them inaccessible through output statistics alone. So, where stealing is no longer feasible, we turn to cloning. We freeze the recovered layers and distill the remaining behavior into the compact clone, trained on public data. This allows the clone to approximate the reasoning of the original model without accessing any of its internal parameters.

\subsection{Stealing Attack: Extracting the Projection Matrix}
\label{sec:logit_stealing}

\begin{figure}[htbp]
  \centering
  \includegraphics[width=1\linewidth]{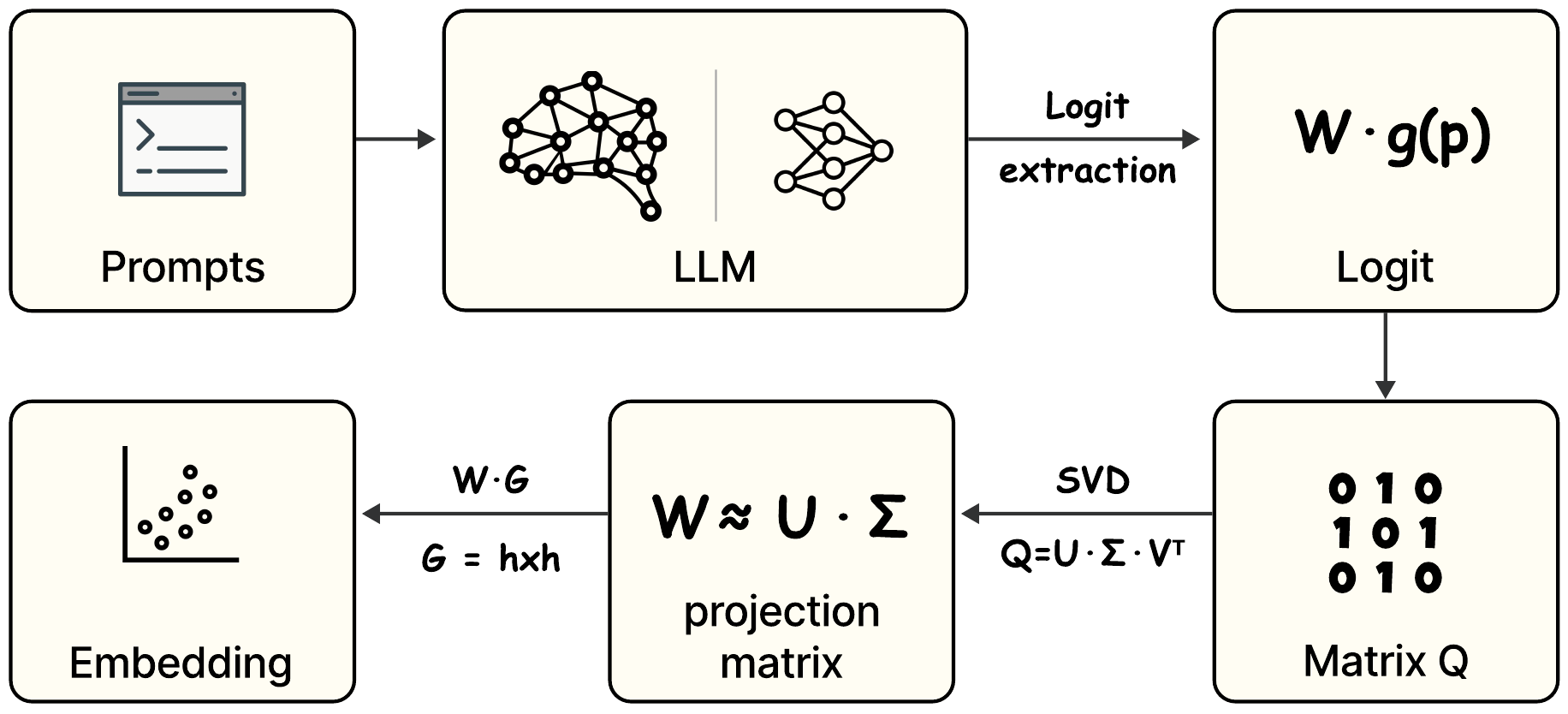}
  \caption{Pipeline of the model stealing attack.}
  \label{fig:model_stealing}
\end{figure}

To recover the output projection matrix $\mathbf{W}$, we query $\mathcal{F}$ with $n>d$ random prompts and collect the top-$k$ logits, as shown in \cref{fig:model_stealing}. Although partial, these responses contain enough information pattern to recover $\mathbf{W}$. We stack these responses into a logit matrix $\mathbf{Q} \in \mathbb{R}^{V \times n}$ and apply SVD:
\begin{equation}
  \mathbf{Q} = \mathbf{U} \mathbf{\Sigma} \mathbf{V}^\top.
\end{equation}

Here, $\mathbf{U}$, $\mathbf{\Sigma}$, and $\mathbf{V}^\top$ are the left singular vectors, singular values, and right singular vectors of $\mathbf{Q}$, respectively. The diagonal entries of $\mathbf{\Sigma}$ tend to drop sharply after the first $d$ components, which helps estimate the hidden size of the model. The top $d$ singular vectors in $\mathbf U$ capture the most important directions in the logit space. By keeping only these top components, we build an estimate of the projection matrix:
\begin{equation}
  \hat{\mathbf W} = \mathbf U_{:,1:d} \, \mathbf \Sigma_{1:d,1:d}.
\end{equation}
This estimate does not exactly match the true projection matrix $\mathbf W$, but it lies in the same column space. In other words, there exists some unknown invertible matrix $\mathbf G\in\mathbb R^{d\times d}$ such that $\hat{\mathbf W} \approx \mathbf{W G}$. Since many downstream tasks only rely on this subspace and $\mathbf G$ is not necessary, having $\hat{\mathbf W}$ is enough to continue the attack effectively. \cref{alg:extraction} provides detailed pseudocode of the described approach.

\begin{algorithm}[htbp]
\caption{Logit-matrix extraction}
\label{alg:extraction}
\begin{algorithmic}[1]
\Require{Target LLM $\mathcal F$, number of queries $n$}
\State sample $\mathbf p_1,\dots,\mathbf p_n$
\State $\mathbf Q\gets[\;]$
\For{$i=1$ to $n$}
  \State $\mathbf Q\gets[\;\mathbf Q\;\;\mathcal F.\texttt{logits}(\mathbf p_i)]$
\EndFor
\State $(\mathbf U,\mathbf \Sigma,\mathbf V)\gets\text{SVD}(\mathbf Q,\text{econ}=true)$
\State $d\gets\arg\max_j\bigl(\log\mathbf \Sigma_{jj}-\log \mathbf\Sigma_{j+1,j+1}\bigr)$
\State $\hat{\mathbf W}\gets \mathbf U_{:,1:d}\,\mathbf \Sigma_{1:d,1:d}$
\State \Return $\hat{ \mathbf{W}}$
\end{algorithmic}
\end{algorithm}


This method assumes access to full or top-k logits, which has been shown to be feasible in commercial black-box APIs through prompt manipulation  \cite{carlini2024stealing}. While quantized or truncated outputs may reduce reconstruction fidelity, the singular value spectrum typically exhibits a sharp drop after the true hidden dimension, enabling reliable estimation of model depth. This recovered projection matrix forms the foundation for the second stage of our pipeline: distilling the internal reasoning behavior into a compact student model.

\subsection{Cloning Attack: Knowledge–Distillation Clone}
\label{subsec:kd}
\begin{figure}[htbp]
  \centering
  \includegraphics[width=1\linewidth]{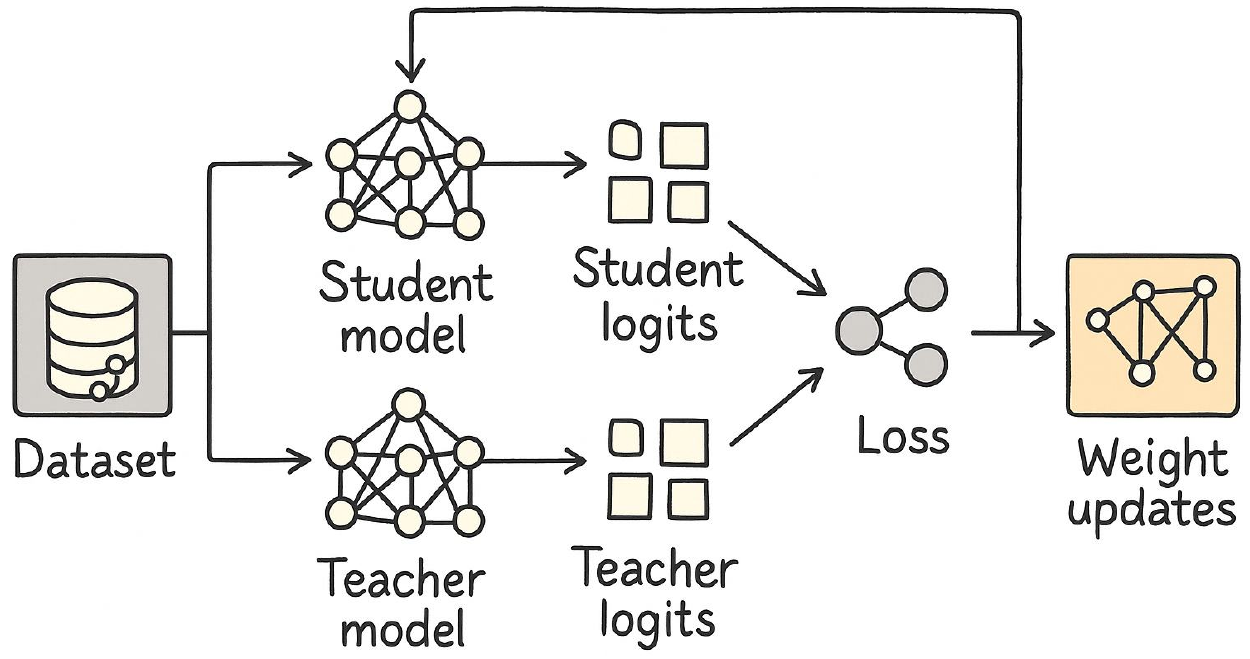}
  \caption{Pipeline of the model cloning attack.}
  \label{fig:knowledge_distilation}
\end{figure}

Once the projection layer is stolen, the remaining transformer blocks remain hidden behind non-linear operations and cannot be accessed directly. Instead of trying to extract them, we craft a family of clone models, commonly referred to as student model in knowledge distillation, to mimic the behavior of the target (a.k.a teacher model). As shown in \cref{fig:knowledge_distilation}, both the target and the student receive the same input, and the student learns to match the target's output via a distillation loss~\cite{hinton2015distilling}.

We sample prompts from open source datasets and use them to query the target model. For each prompt $\mathbf{p}$, the target produces logits $z_T = f_T(\mathbf{p})$ and the clone produces $z_S = f_S(\mathbf{p})$. These logits are softened by temperature scaling:
\begin{equation}
\mathbf{s}_T = \mathrm{softmax}\left(\frac{z_T}{\tau}\right), \quad \mathbf{s}_S = \mathrm{softmax}\left(\frac{z_S}{\tau}\right)
\end{equation}

Here, $\mathbf{s}_T$ and $\mathbf{s}_S$ represent the predicted token probabilities from the target and the student (clone), respectively, and $\tau$ is the temperature hyperparameter used to control the sharpness of the distribution. The training loss compares the two outputs and adds a small cross-entropy term with the true label $y$:
\begin{equation}
\label{eq:kd-loss}
\mathcal{L}(\mathbf{p}) = \tau^2 \, \mathrm{KL}(\mathbf{s}_T \parallel \mathbf{s}_S) + \lambda\, \mathrm{CE}(z_S, y)
\end{equation}
where $\mathrm{KL}(\cdot \parallel \cdot)$ is the KL divergence between the softened teacher and student outputs, $\mathrm{CE}(\cdot,\cdot)$ is the standard cross-entropy loss with the original ground-truth labels, and $y$ is the true label from the dataset. In our setting, the KL loss is prioritized by setting $\lambda=0.1$ so that the student mainly focuses on matching the teacher's behavior.

\begin{table*}[htbp]
  \centering
  \caption{Fidelity and efficiency of cloned models on WikiText-2}
  \label{tab:WikiText2_clone_quality}
  \begin{tabular}{lccccccccccc}
    \hline
    Model & Depth & Params & NLL $\downarrow$ & PPL $\downarrow$ & $\Delta$PPL (\%) $\downarrow$ &
           KL $\downarrow$ & Cos-sim $\uparrow$ & Memoriz. $\downarrow$ & Mem-match $\downarrow$ & Speed up $\uparrow$ & \makecell{Size \\Reduction $\uparrow$} \\
    \hline
    Student-4 & 4 & 67.0\,M & 7.608 & 2013
    & 10.27\% & 19.26 & 0.9730 & High & 1.8\% & \textbf{17.1\%} & \textbf{18.1\%} \\
    Student-5 & 5 & 74.1\,M & 7.592 & 1982
    & 8.57\% & 17.32 & 0.9741 & High & 1.4\% & 5.9\%  & 9.5\% \\
    Student-6 & 6 & 81.2\,M & 7.580 & 1959
    & 7.31\% & 15.66 & \textbf{0.9765} & Mid & 1.1\% & -3.5\% & -0.8\% \\
    Student-7 & 7 & 88.3\,M & \textbf{7.562} & \textbf{1922
    } & \textbf{5.30}\% & 14.37 & 0.9750 & \textbf{Low} & \textbf{0.5\%} & -13.7\% & -7.8\% \\
    Student-8 & 8 & 95.4\,M & 7.571 & 1940
    & 6.25\% & \textbf{13.92} & 0.9760 & Mid & 1.1\% & -20.4\% & -16.5\% \\
    \hline
    Teacher  & 6 & 81.0\,M & 7.510 & 1826
    & 0.00\% & 0.00 & 1.000 & Baseline & - & - & - \\
    \hline
  \end{tabular}
\end{table*}

\begin{figure*}[htbp]
\centering
\includegraphics[width=\linewidth]{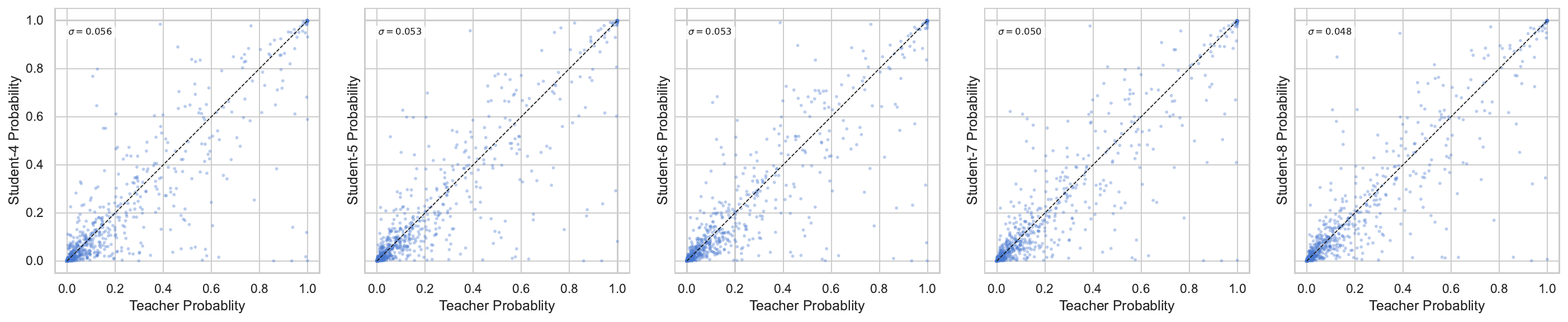}
\caption{Token-probability alignment between each clone and the target.}
\label{fig:prob_align}
\end{figure*}

The clone keeps the stolen embedding and projection layers fixed. Only the transformer blocks are trained. To explore how depth affects learning, we try different versions of the clone with 4 to 8 layers. The full process is shown in~\cref{alg:kd}.

\begin{algorithm}[htbp]
\caption{Knowledge–Distillation Clone}
\label{alg:kd}
\begin{algorithmic}[1]
\Require Prompt set $\mathcal{P}$, teacher model $f_T$, student model $f_S$
\State Initialize $f_S$ 
\For{epoch = 1 to $E$}
    \For{minibatch $\mathcal{B} \subset \mathcal{P}$}
        \State Update temperature $\tau$ based on epoch
        \ForAll{prompt $\mathbf p \in \mathcal{B}$}
            \State Compute $\mathbf{s}_T = \mathrm{softmax}(f_T(\mathbf p)/\tau)$
            \State Compute $\mathbf{s}_S = \mathrm{softmax}(f_S(\mathbf p)/\tau)$
            \State Compute $\mathcal{L}(\mathbf p)$ using \cref{eq:kd-loss}
        \EndFor
        \State Update $f_S$ by minimizing the average loss over $\mathcal{B}$
    \EndFor
\EndFor
\State \Return trained student model $f_S$
\end{algorithmic}
\end{algorithm}

This distillation process enables the student model to approximate the target’s internal reasoning behavior using only black-box access and public data. By freezing the recovered projection layer and training compact transformer variants, we produce deployable clones that preserve both output fidelity and latent geometry, without ever accessing the target’s internal parameters. We evaluate the effectiveness of this approach in the following section.

\section{Experimental Verification}
\label{sec:experimental_analysis}
In this section, we evaluate our steal-and-clone pipeline using  distilGPT-2, a black-box LLM with 6 transformer layers and $\approx81$M  parameters. The objective is to replicate its behavior into student models with 4 to 8 transformer layers, trained solely on public prompts and black-box access. We sample $\mathcal{P}$ in \cref{alg:kd} from WikiText-2~\cite{mettu2022wikitext2}, and assess performance on both WikiText-2 (in-distribution) and WikiText-103~\cite{dekomposition2022wikitext103} (out-of-distribution) to evaluate generalization ability.

\subsection{In-Distribution Fidelity and Efficiency}
\cref{tab:WikiText2_clone_quality} compares the performance of the cloned models on WikiText-2 using multiple fidelity and efficiency metrics. We use negative log-likelihood (NLL) and perplexity (PPL) to evaluate language modeling quality~\cite{nasr2023scalable}, where lower values reflect better predictions. The relative perplexity increase (PPL \%) shows how far each clone deviates from the target. KL divergence~\cite{sunoj2025quantile} measures the difference in token distributions between the target and the clone and cosine similarity tracks how closely their internal representations align. We also present qualitative indicators of memorization behavior, how often the model repeats seen data, and memory-match error, which captures the percentage of mismatched internal memory states. Finally, we include inference speedup and size reduction relative to the target to assess practical deployability.

The 6-layer clone (Student-6) achieves the best overall balance across these dimensions, with only a 7.3\% perplexity gap, low KL divergence (15.66), and the highest cosine similarity 97.65\%, closely matching the teacher's behavior while  maintaining comparable size and speed. Student-7 offers the best raw fidelity in terms of perplexity and memory alignment but incurs increased computational cost. On the efficiency side, Student-4 delivers the strongest gains with a 17.1\% speedup and 18.1\% size reduction, while still staying within 10.3\% of the teacher’s perplexity and retaining strong alignment.

These results demonstrate that our cloning method can recover and compress models that are both compact and highly faithful to the target’s behavior, offering flexibility depending on the trade-offs needed for a specific deployment.


\subsection{Token-Level Output Alignment}
\cref{fig:prob_align} shows how closely each cloned model matches the teacher's output probabilities for individual tokens. Each dot represents a token, and proximity to the diagonal line indicates stronger agreement between the student and teacher. Student-6 aligns particularly well with the teacher, with a variance ($\sigma = 0.053$), comparable to smaller models Student-5 ($\sigma = 0.053$) and better than Student-4 ($\sigma = 0.056$). This means it has learned to mimic the teacher very efficiently. Larger models (Student-7 and Student-8) show even tighter alignment ($\sigma = 0.050$ and $0.048$), but Student-6 achieves this fidelity with the same parameter count as the teacher. These results are consistent with the KL divergence trends reported in \cref{tab:WikiText2_clone_quality} and reinforce the effectiveness of our distillation strategy.

\subsection{Generative Output Diversity and Memorization}
\begin{figure}[htbp]
\centering
\includegraphics[width=\linewidth]{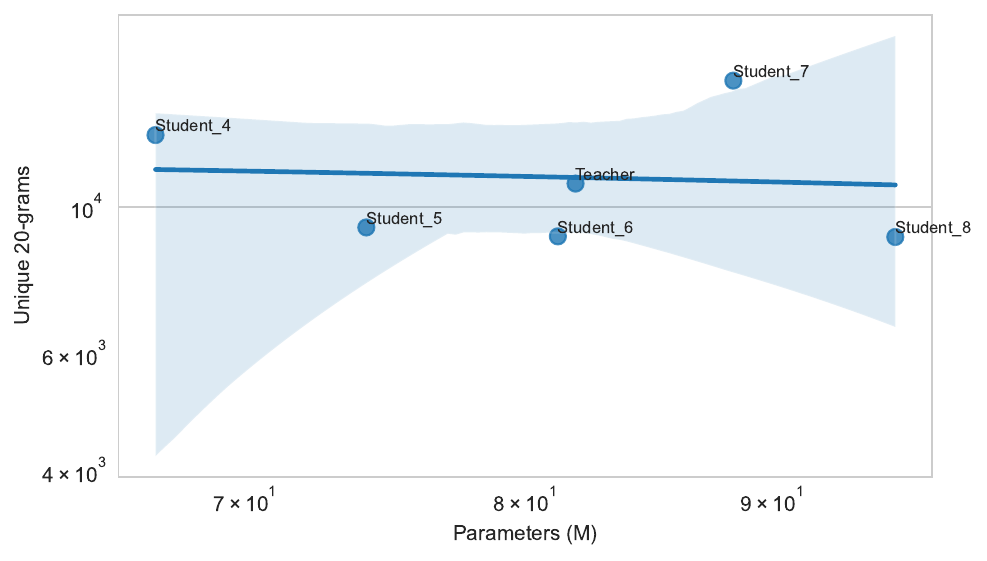}
\caption{Unique 20-grams as a function of parameter count.}
\label{fig:param_mem}
\end{figure}

We measure the number of unique 20-grams, distinct sequences of 20 consecutive tokens, produced by each model to assess generative diversity.  A higher number means the model produces novel text, while a lower number may suggest it repeats seen patterns or memorizes training data. 
The blue line in \cref{fig:param_mem} indicates the overall trend, with the shaded area showing expected variation. Student-4 and Student-7 exceed the teacher’s diversity, suggesting stronger generalization, while Student-5 and Student-6 fall below, likely due to conservative decoding or underfitting. Interestingly, Student-8, despite its size, generates fewer unique sequences, indicating that increased capacity does not guarantee improved diversity.

\subsection{Model Selection Criteria}
\begin{figure}[htbp]
    \centering
    \includegraphics[width=\linewidth]{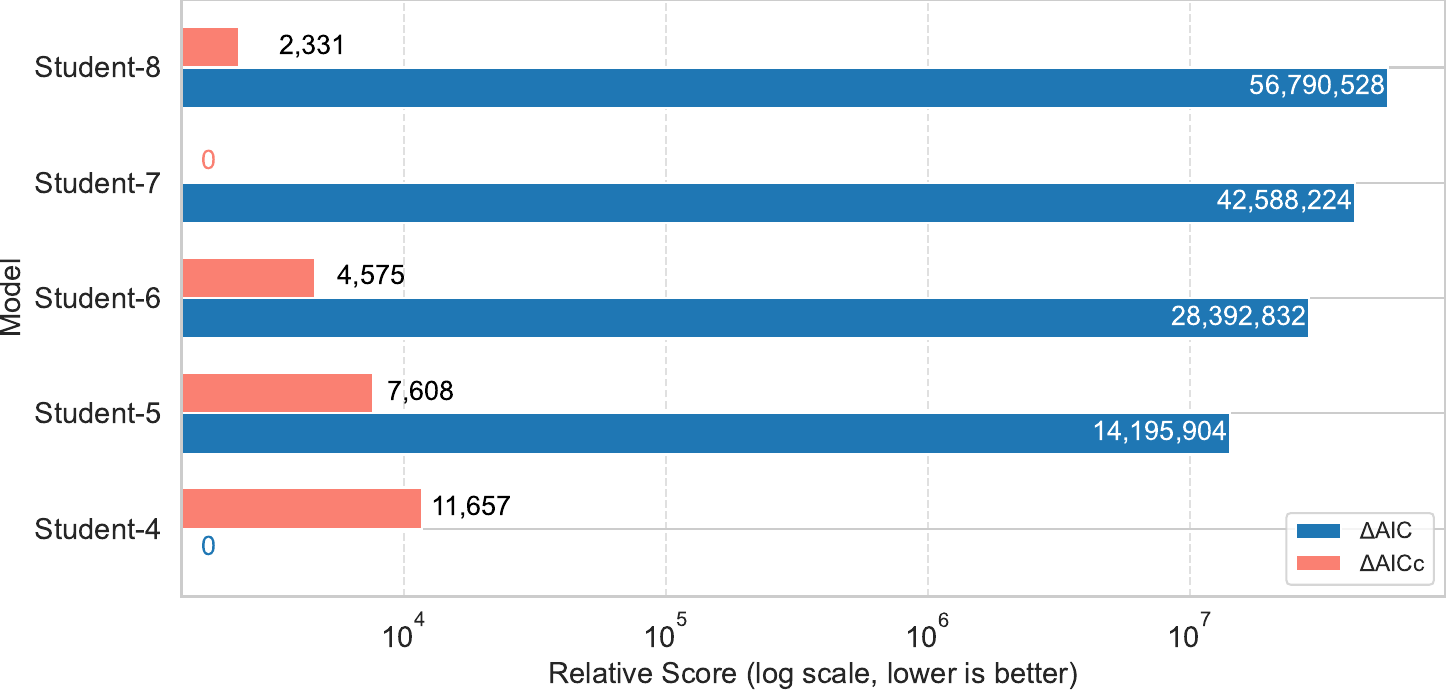}
    \caption{Relative AIC and AICc scores of cloned models.}
    \label{fig:delta_aic}
\end{figure}

We apply Akaike Information Criterion ($\Delta$AIC) and its corrected variant for small sample sizes ($\Delta$AICc) to evaluate model fit while penalizing complexity, with lower values indicating better trade-offs.
\cref{fig:delta_aic} shows the difference from the best model, which is assigned zero. Student-7 achieves the lowest AIC, indicating the best raw fit. However, Student-4 performs best under AICc, suggesting superior generalization when accounting for model size and limited data. The widening gap between AIC and AICc for larger models highlights the diminishing returns of added parameters.

\subsection{Generalization to Unseen Data}
The results discussed so far are based on the WikiText-2 dataset, which was also used during the knowledge distillation process. Since the clones were fine-tuned on this data, it is possible that they learned not just the target model's behavior but also memorized patterns specific to that dataset. To test generalization, we evaluate all models on WikiText-103~\cite{dekomposition2022wikitext103}, a larger and more diverse dataset not used during training.

\cref{tab:wt103_results} demonstrates all clones maintain strong performance, with perplexity gaps under 10.5\%. Student-6 again offers the best trade-off, with only a 7.7\% increase in perplexity. KL divergence trends mirror those from WikiText-2, with deeper models better aligning to the target. These results confirm that the clones 
capture the target model’s latent reasoning rather than memorizing training prompts.

\begin{table}[htbp]
\centering
\caption{Cloned model performance on WikiText-103 (unseen prompts)}
\label{tab:wt103_results}
\begin{tabular}{lccccc}
\hline
Model & Depth & NLL $\downarrow$ & Perplexity $\downarrow$ & PPL (\%) $\downarrow$ & KL $\downarrow$ \\
\hline
Student-4 & 4 & 7.616 & 2029.67 & 10.5\% & 18.99 \\
Student-5 & 5 & 7.598 & 1993.30 & 8.5\% & 17.26 \\
Student-6 & 6 & 7.590 & 1977.62 & 7.7\% & 15.58 \\
Student-7 & 7 & \textbf{7.570} & \textbf{1938.95} & \textbf{5.6\%} & 14.38 \\
Student-8 & 8 & 7.576 & 1950.25 & 6.2\% & \textbf{13.84} \\
\hline
Target    & 6 & 7.516 & 1837.11 & 0.0\% & 0.00 \\
\hline
\end{tabular}
\end{table}

\subsection{Parameter Efficiency}
\begin{figure}[htbp]
    \centering
    \includegraphics[width=\linewidth]{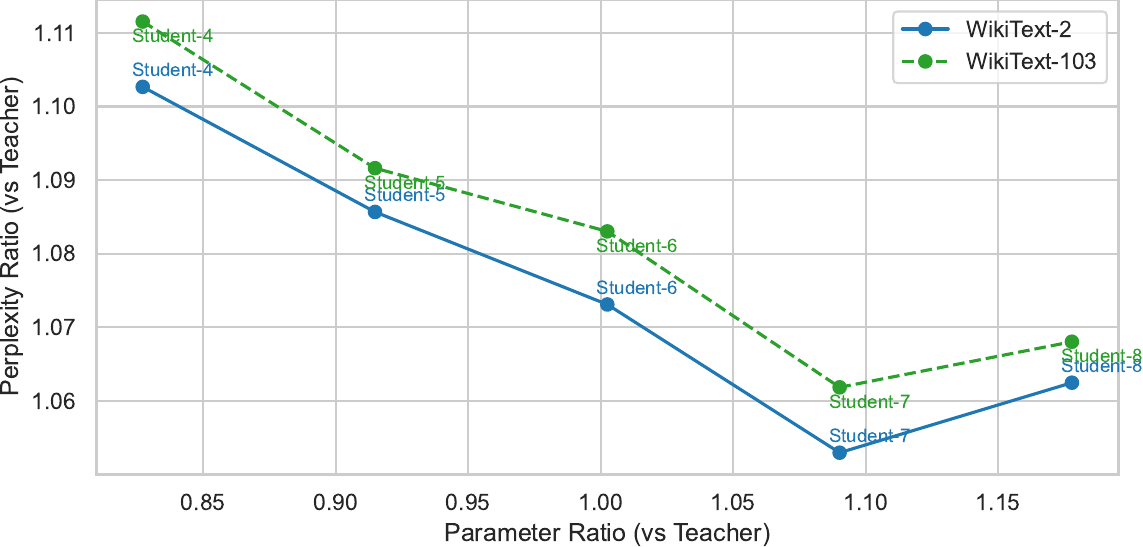}
    \caption{Normalized perplexity vs model size on WikiText-2 and WikiText-103. Both axes are scaled relative to the teacher.}
    \label{fig:norm_ppl_vs_size}
\end{figure}

\cref{fig:norm_ppl_vs_size} shows the normalized perplexity versus the normalized parameter count to better understand the trade-off between model size and performance. The x-axis shows the relative size of each model, while the y-axis shows how much their perplexity differs from the target. We see that all cloned models remain close to the teacher in terms of performance. Among them, Student-7 gives the best overall trade-off, it is slightly larger than the target but achieves even lower perplexity. Student-6 matches the target in size and performs nearly as well. As expected, smaller models like Student-4 and Student-5 show modest degradation, consistent with reduced capacity.

\begin{figure}[htbp]
    \centering
    \includegraphics[width=\linewidth]{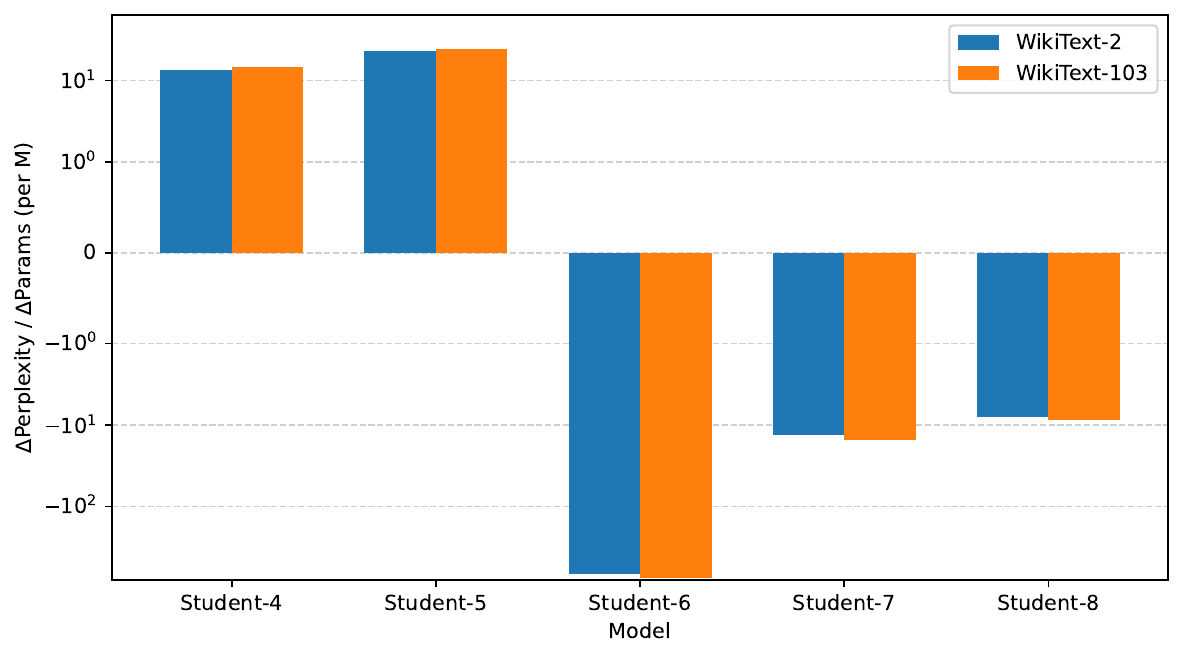}
    \caption{Perplexity increase per million parameters saved compared to the teacher. Lower bars mean better trade-off.}
    \label{fig:ppl_efficiency}
\end{figure}

\cref{tab:ppl_efficiency} further analyzes how efficient each model is by measuring the change in perplexity per million parameters. Most models show small changes, but Student-6 stands out with a negative cost, with a sharp drop in perplexity despite using the same number of parameters as the target. This indicates that the cloned Student-6 model uses its capacity more efficiently, likely due to good alignment with the recovered weights. Overall, the results highlight that cloned models remain competitive, and in some cases even outperform the target, while being more efficient in parameter usage.
\cref{fig:ppl_efficiency} also supports same observation by visualizing the trade-off between perplexity and model size in log-scale.

\begin{table}[htbp]
\centering
\caption{Per–parameter perplexity cost when cloning the target model
}
\label{tab:ppl_efficiency}
\begin{tabular}{lcccc}
\hline
\multicolumn{1}{c}{\shortstack{Cloned \\ Model}} &
\multicolumn{1}{c}{\shortstack{$\Delta$PPL\\ \textit{(WT‑2)}}} &
\multicolumn{1}{c}{\shortstack{$\Delta$PPL\\ \textit{(WT‑103)}}} &
\multicolumn{1}{c}{\shortstack{$\dfrac{\Delta \mathrm{PPL}}{\Delta\mathrm{Param}}$\\ \textit{(WT‑2)}}} &
\multicolumn{1}{c}{\shortstack{$\dfrac{\Delta \mathrm{PPL}}{\Delta\mathrm{Param}}$\\ \textit{(WT‑103)}}} \\
\hline
Student‑4 & 187.47 & 203.67 & \phantom{$-$}13.39 & \phantom{$-$}14.55 \\
Student‑5 & 156.44 & 167.30 & \phantom{$-$}22.67 & \phantom{$-$}24.25 \\
Student‑6 & 133.50 & 151.62 & \textbf{$-$667.48} & \textbf{$-$758.09} \\
Student‑7 &  96.73 & 112.95 &  $-$13.25 &  $-$15.47 \\
Student‑8 & 114.09 & 124.25 &  $-$7.92 &  $-$8.63 \\
\hline
\end{tabular}
\end{table}

These experiments confirm that our cloned models maintain high fidelity to the target across both in-distribution (WikiText-2) and out-of-distribution (WikiText-103) datasets. Student-6 offers the best trade-off between fidelity and efficiency, while Student-4 demonstrates that significant compression is possible with minimal performance degradation. These findings validate the effectiveness of our distillation strategy under constrained query budgets.

\section{Conclusion}
\label{sec:conclusion}

This work exposes a critical vulnerability in the deployment of LLMs via inference APIs that expose top-$k$ logits. We show that even under constrained query budgets and limited computational resources, an adversary can reconstruct a high-fidelity replica of a production-grade LLM using only black-box access. Our two-stage pipeline comprising projection matrix recovery via SVD and task-aware knowledge distillation, enables the creation of compact student models that preserve both output fidelity and internal representation geometry.

Empirical results show that a 6-layer student model achieves 97.6\% cosine similarity with the teacher model and maintains a perplexity gap of just 7.3\%, while a 4-layer variant offers significant gains in inference speed and model size with minimal degradation. These findings confirm that partial logit leakage can be transformed into a deployable clone capable of generalizing to unseen data, underscoring the operational risks posed by unsecured inference endpoints.

Beyond the technical contributions, this work highlights the broader ethical and governance implications of black-box LLM replication. The ability to reconstruct high-fidelity model clones raises urgent concerns around the circumvention of alignment safeguards \cite{sun2024multi}, unauthorized redistribution of proprietary systems \cite{carlini2024stealing}, and potential leakage of memorized sensitive data \cite{panda2024teach,nasr2023scalable}. These risks underscore the need for responsible disclosure practices and the development of multi-layered defense strategies that integrate technical, operational, and legal safeguards\cite{oliynyk2023know,irtiza2024llm,mazeika2024harmbench}.

We encourage future research to build upon this foundation by exploring robust watermarking schemes \cite{oliynyk2023know}, privacy-preserving inference protocols \cite{feng2024privacy}, and policy frameworks that can deter misuse while preserving the utility of open-access AI. By surfacing these considerations, we aim to position this work as a reference point for interdisciplinary efforts that seek to secure the deployment of LLMs in high-stakes environments.

Future work will extend this threat model to APIs that expose only top-$k$ probabilities or quantized outputs, and to multi-modal systems that integrate text with other modalities such as code, imagery, and audio. We also plan to evaluate the effectiveness of countermeasures including adaptive noise injection, behavioral fingerprinting, and secure on-premise inference to mitigate emerging model exfiltration threats.

\section*{Acknowledgment}
Portions of this manuscript were augmented with the assistance of Microsoft 365 Copilot Researcher and Writing Coach Agents (Microsoft, 2025). The final content was reviewed and confirmed by the authors.

\bibliographystyle{IEEEtran}
\bibliography{IEEEabrv,references}

\end{document}